# Analysing Word Importance for Image Annotation

Payal Gulati, A. K. Sharma

Computer Engineering Department, YMCA UST
Faridabad, Haryana, India

**Abstract**
Image annotation provides several keywords automatically for a given image based on various tags to describe its contents which is useful in Image retrieval. Various researchers are working on text based and content based image annotations [7,9]. It is seen, in traditional Image annotation approaches, annotation words are treated equally without considering the importance of each word in real world. In context of this, in this work, images are annotated with keywords based on their frequency count and word correlation. Moreover this work proposes an approach to compute importance score of candidate keywords, having same frequency count.

## 1. Introduction

The importance of image annotation has been increasing with the growth of the worldwide web. Finding relevant digital images from the web and other databases is a difficult task because many of these images do not have relevant annotations [4,5]. There are various methods to deal with image annotations but it is observed that these approaches [6,7,9] treat each candidate word equally. In the light of this, in this work, words are associated with importance score instead of being considered equally. Annotation words should be associated with importance measurement, instead of being considered equally. In this work, a method to compute an importance scores for every candidate keyword is proposed. In this approach, collaboration of word frequency and word correlation is used for annotating images. Moreover importance score of keywords with same frequency is determined, for instance, suppose $word_1$ and $word_2$ have the same frequency count, but $word_1$ is correlated with $word_3$ which has higher importance as compared with $word_2$ which is correlated with $word_4$ having low importance; thus $word_1$ will be considered important as compared to $word_2$.

This paper is organised in the following way. Section 2 discusses the relevant work done in this domain. Section 3 presents the proposed architecture. Algorithm and implementation results are discussed in section 4 and 5 respectively. Finally Section 6 comprises of the conclusion.

## 2. Related Work

Keywords are most important Search Engine Optimizing [1,8] element as they are the search strings that are matched against. Therefore importance of keyword needs to be analyzed before annotating and indexing them. Some researchers have worked on analysis of keyword on the basis of keyword density [2,3]. The general idea is higher the keyword density, more relevant to the search string the page is. Some work is done on tracking the keyword on special places which proves that keywords are not only selected on the basis of their quantity i.e. frequency but on quality as well. For instance keywords in page title, headings are of more importance compared to the other text. In the light of this, this work focuses on calculating the importance of keyword on the basis of word correlation irrespective of their same frequency/occurrence count.

## 3. Proposed Architecture

The proposed architecture shown in Fig. 1 is discussed in this section.

### 3.1 WWW and Web Page
WWW is a vast repository of web pages. Crawl manager takes the seed URL from the URL list and fetches the page from www.

### 3.2 Parser
Parser parses the web page fetched by crawl manager from www. It then parses for tags such as image tags, alt text, metadata, page title tags etc. This phase returns all the keywords fetched during parsing to the frequency calculator.

### 3.3 Frequency Calculator
In this phase, frequency count of all the keywords returned by the parser is determined. Candidate keywords are further determined based on the threshold value.

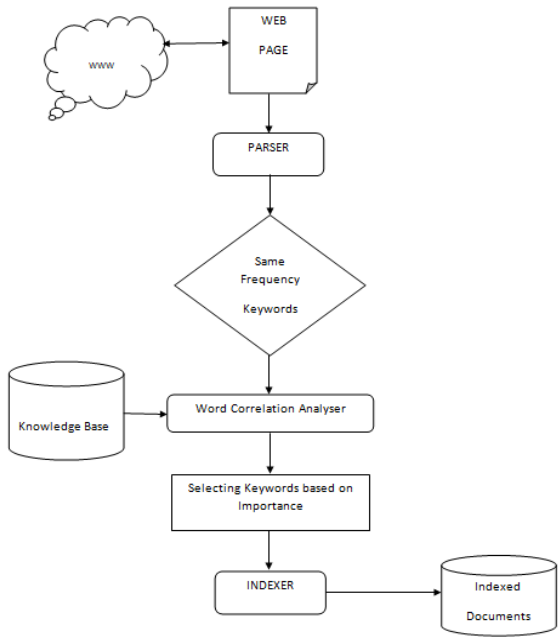

Fig. 1 Proposed Architecture

### 3.4 Same Frequency Keyword Detector
Same frequency keyword detector detects the candidate keywords having same frequency count.

### 3.5 Knowledge base
The knowledge base is a repository of extracted rules which have been derived using the association rule mining. A knowledge base containing rules is shown in Fig.2.

| ID | TermsX | TermsY |
|---|---|---|
| 3 | water | mountain |
| 4 | himalaya | mountain |
| 5 | nature | mountain |
| 6 | tree | nature |
| 7 | leaves | tree |
| 8 | greenary | leaves |
| 9 | bird | nature |
| 10 | greenary | grass |
| 11 | nature | sky |
| 12 | sun | sky |

Fig. 2 Knowledge Base

### 3.6 Word Correlation Analyser and Important Keyword Selector
In this module correlation between words is determined based on association knowledge base. In this module, keyword importance is analysed based on word correlation. Suppose word w1 and w2 have the same frequency, but w1 was correlated to word w3 which has higher importance, while w2 was correlated with w4 which has lower importance. Thus w1 will be considered important compared to w2.

### 3.7 Indexer
This modules indexes www images along with their keywords in indexed repository. Without an index, the search engine would scan every document in the corpus for the queried image, which would require considerable time and computing power.

### 3.8 Indexed Repository
Image Search engine component crawler collects, parses, and stores images in an indexed repository to facilitate fast and accurate image retrieval.

## 4. Algorithm

ImportanceScoreCalculation
Input : Same Frequency Keywords
Output: Important Keyword

---
*Algorithm ImportanceScoreCalculation*
*Begin*
*for every keyword with same frequency count*
*calculate the Rank of keyword using*
$R(b) = \sum_{b \in CB} (T(a,b) + R(a))$ *where a is the backword of b i.e.*
$a \rightarrow b$, *R(b) is the rank of word b that belong to set B that contains all the forward words in the knowledge base, R(a) is the rank of a.*
*If $a \rightarrow b$ (a is a back word of b)*
    $T(a,b) = 1/Ntotal$ *where Ntotal is the total no. of keywords in the knowledge base*
*else*
    $T(a,b) = 0$
*end*

---

Let A,B,C,D,E,F,G,H,I be keywords and their corresponding frequencies are shown in Table 1.

Table 1. Keywords and their Frequency

| Keywords | Frequency Count |
|---|---|
| A | 4 |
| B | 6 |
| C | 3 |
| D | 6 |
| E | 5 |
| F | 2 |
| G | 2 |
| H | 1 |
| I | 3 |
| J | 1 |

Considering 40% threshold value, Candidate keywords obtained from Table 1 is shown in Table 2.

Table 2. Candidate keywords

| Candidate Keywords | Frequency Count |
|---|---|
| A | 4 |
| E | 5 |
| B | 6 |
| D | 6 |

Importance score for same frequency keywords B and D can be calculated using the proposed algorithm ImportanceScoreCalculation.

Table 3. below show the word correlations

Table 3. Word Correlation

| Word Correlations |
|---|
| a → b |
| c → b |
| d → b |
| e → d |
| f → e |
| g → f |
| h → d |
| g → j |

According to the algorithm, Rank of B and D is calculated

$N_{Total}$ = 1/10 = 0.1

R(b) = [0.1 + R(a)] + [0.1 + R(c)] + [0.1 + R(d)]
    = 0.1 + 0.1 + [0.1 + [[0.1 + R(e)] +[0.1 +R(h)]]
    = 0.1 + 0.1 + [0.1 + 0.1 + [0.1 + R(f)] + 0.1
    = 0.1 + 0.1 + 0.1 + 0.1 + 0.1 + [0.1 + R(g)] + 0.1
    = 0.1 + 0.1 + 0.1 + 0.1 + 0.1 + 0.1 + 0.1
    =0.7         (1)

R(d) = [0.1 + R(e)] + [0.1 + R(h)]
    = [0.1 + [0.1 + R(f)] + 0.1
    = 0.1 + 0.1 + [0.1 + R(g)] + 0.1
    = 0.1 + 0.1 + 0.1 + 0.1
    = 0.4         (2)

From (1) and (2) it is seen that keyword b is more important compared to keyword d as per their correlation score.

## 5. Implementation

To compute the importance score of candidate keywords, the proposed algorithm has been implemented in java. The implemented result for the same is shown in Fig. 3.

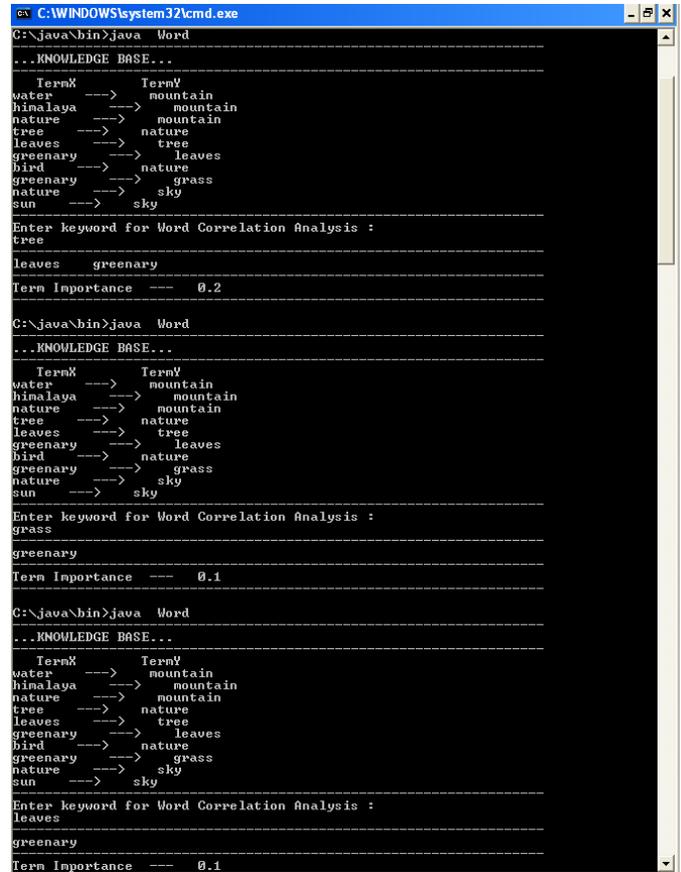

Fig. 3 Implementation of the proposed approach

## 6. Conclusion

This work proposes an approach for analysing word importance for image annotation based on frequency count and word correlation information. An approach proves to be efficient, as important keywords are analysed for annotating images.